\documentclass[12pt]{article}
\usepackage{amsfonts}
\usepackage{latexsym}
\begin{document}
\begin{titlepage}
\rm
\normalsize
\begin{flushright} 
UTTG-02-00

\texttt{hep-th/0001073}
\end{flushright} 
\vskip 2.5cm
\begin{center}
{\Large \bf Dirac Monopole in Non-Commutative Space}
\vskip 1.5cm
Li Jiang
\vskip 0.5cm
\textit{Theory Group, Department of Physics\\
        University of Texas at Austin\\
        Austin, TX 78712, USA\\}
\texttt{jiang@physics.utexas.edu}
\vskip 3.5cm
\textbf{Abstract}
\end{center}
\vskip 0.5cm
\noindent We consider static $U(1)$ monopole in non-commutative space.
Up to the second order in the non-commutativity scale $\theta$, we
f\mbox{}ind no non-trivial corrections to the Dirac solution, the monopole
mass remains inf\mbox{}inite. We argue the
same holds for any arbitrary higher order. Some speculation about the
nature of non-commutative spacetime and its relation to the cosmological
constant is made.  
\end{titlepage}

\noindent Non-commutative geometry arises naturally in string theory
when the Neveu-Schwarz $B$ f\mbox{}ield is turned on. In a certain limit,
the low-energy ef\mbox{}fective theory of the worldvolume  is described 
by gauge theory in 
non-commutative space~\cite{sw}. Motivated by the attempt to investigate
localized structures, some researchers have recently studied topological 
f\mbox{}ield conf\mbox{}iguration
in non-commutative geometry, namely the non-Abelian monopoles~\cite{hh,hhm,db}.
In this letter, we consider another elementary example, the Dirac 
monopole~\cite{pd}.

It would be helpful to make some comparisons f\mbox{}irst. 
Non-Abelian monopole 
can be visualized as D-string stretched between 
branes~\cite{dl,dd}. In the $U(2)$ case, when a background $B$ f\mbox{}ield 
is turned on along the branes, the string is tilted because the two
endpoints carry opposite charges~\cite{hh}. This leads to a dipole structure
in the magnetic f\mbox{}ield of the monopole. Explicit calculation of
this ef\mbox{}fect to the $O(\theta)$ order has been carried out 
in~\cite{hhm,db}.
However, $U(1)$ monopole does not admit such a simple geometric picture.
First of all, there is no need to introduce a Higgs f\mbox{}ield $\phi$. The
singularity is put in by hand: one simply adds a source term to the Bianchi
identity, then the base manifold becomes $\mathbb{R}^3\setminus \{0\}$, which
deformation-retracts to $S^2$. The Wu-Yang method~\cite{wy} is applied
to yield the quantization of magnetic monopole charge 
($\pi _1(U(1))=\mathbb{Z}$). By contrast the topological invariant of a
non-Abelian monopole is def\mbox{}ined by the asymptotic behavior of the Higgs
f\mbox{}ield. Second, the energy of Dirac monopole diverges, while that of 
't~Hooft-Polyakov monopole~\cite{gh,ap} is f\mbox{}inite ($m\propto 
1/g_{\rm YM}$). 
Although it's dif\mbox{}f\mbox{}icult 
to make topological argument in non-commutative space, it has been shown
that the BPS bound~\cite{eb,ps} still exists~\cite{hh,hhm,db}. One naturally 
asks whether non-commutativity will render Dirac monopole a f\mbox{}inite mass.
This is interesting since all the attempts to f\mbox{}ind magnetic 
monopole have
failed. We can even pose the question: why should one treat a ``particle''
with inf\mbox{}inite mass as a physical entity? The case should be compared
to that of electron. Although the f\mbox{}ield energy due to a point 
electric charge diverges in the same way, it's renormalizable as electron
has an experimentally measurable mass about half MeV (one simple way to
get rid of the inf\mbox{}inity is to replace the point source with a smooth
compactly-supported charge, while the same procedure is not applicable
to monopole~\cite{jt}). 

If Dirac monopole became f\mbox{}initely heavy in
non-commutative space, it would be of great interests to theorists and
experimentalists alike. However, our calculation gives a negative answer. 
Up to the $O(\theta ^2)$ order,
we show explicitly that there is no correction to the $U(1)$ gauge 
connection $A$ (except for $\theta$-dependent gauge transformation), 
therefore the mass remains inf\mbox{}inite. We
argue the same is true for any higher order. In the following, we will present
our calculation by adopting a mixed notation of tensor and dif\mbox{}ferential
form. We also discuss another formulation of the $U(1)$ monopole in
non-commutative space. Finally, we ponder over the nature of non-commutative
spacetime and its relation to the cosmological constant.

The f\mbox{}ield strength in non-commutative gauge theory is 
def\mbox{}ined as usual 
with replacement of the ordinary product by the $\star$-product\footnote{
$f(x)\star g(x)=\exp (\frac{i}{2}\theta_{ij}\partial_i\partial_j')
               f(x)g(x')|_{x=x'}$ and $[x_i, x_j]=i\theta_{ij}$.}
\begin{equation}
F=dA-\frac{i}{2}[A, A]_{\star}.
\end{equation}       
Up to the second order $O(\theta ^2)$, $F$ in component form 
is\footnote{We only consider static situation and set $A_0=0$ since there
is no electric source.}
\begin{equation} \label{eq:F_ij}
F_{ij}\simeq \partial_i A_j-\partial_j A_i+\theta_{mn}\partial_m A_i
                                                \partial_n A_j,
\end{equation}    
where $i$, $j$ run from 1 to 3. One notices there is no $O(\theta ^2)$ term.
In fact, all the even powers of $\theta$ vanish because of its antisymmetric
nature. This can also be seen from the def\mbox{}inition of $F$: 
all the even powers
contain an $i$, but $F$ is a real number. Then what do we mean by the second
order correction? Of course presumably the gauge potential $A$ itself should
receive some corrections due to non-commutativity. Expanding $A$ to 
$A\simeq A^0+A^1+A^2$ (the upper indices denote the order in $\theta$), we 
rewrite (\ref{eq:F_ij}) as $F_{ij}\simeq F_{ij}^0+F_{ij}^1+F_{ij}^2$, where
{\setlength\arraycolsep{2pt}
\begin{eqnarray}
F_{ij}^0 & = & \partial_i A_j^0-\partial_j A_i^0 \nonumber\\
F_{ij}^1 & = & \partial_i A_j^1-\partial_j A_i^1+\theta_{mn}\partial_m A_i^0
                                       \partial_n A_j^0   \\
F_{ij}^2 & = & \partial_i A_j^2-\partial_j A_i^2+\theta_{mn}\partial_m A_i^0
    \partial_n A_j^1+\theta_{mn}\partial_m A_i^1 \partial_n A_j^0. \nonumber
\end{eqnarray}}

\vskip -0.5cm
\noindent For later convenience, we def\mbox{}ine $f^{1, 2}=dA^{1, 2}$. 
The goal is 
to f\mbox{}ind the solution to the modif\mbox{}ied Bianchi 
identity (in Gauss units)
\begin{equation} \label{eq:DF}
DF=4\pi g\delta^3(\vec r\,)\ast\!1
\end{equation}
where $g$ is the monopole charge and $\ast1$ is the volume form. The
covariant derivative of $F$ in non-commutative space is similarly 
def\mbox{}ined
\begin{equation}
DF=dF-i[A, F]_{\star}.
\end{equation}
Order by order in $\theta$, (\ref{eq:DF}) is
{\setlength\arraycolsep{2pt}
\begin{eqnarray}
dF^0 & = & 4\pi g\delta^3(\vec r\,)\ast\!1 \nonumber\\
dF^1 & = & -\theta_{mn}\partial_m A^0\wedge \partial_n F^0 \\
dF^2 & = & -\theta_{mn}\partial_m A^1\wedge \partial_n F^0
           -\theta_{mn}\partial_m A^0\wedge \partial_n F^1. \nonumber
\end{eqnarray}}

\vskip -0.5cm
\noindent The zeroth order equation is just the familiar 
$\nabla \cdot \vec B^0=4\pi g\delta^3(\vec r\,)$, with $B^0=\ast F^0$.
Its solution is simply $\vec B^0=g\vec r/r^3$, and $A^0$ has singularities
(Dirac string) when expressed in a single coordinate system. One then 
solves for $A^{1, 2}$ by plugging $A^0$ into the f\mbox{}irst and the 
second order equations. In component form, the f\mbox{}irst order equation
reads
{\setlength\arraycolsep{2pt}
\begin{eqnarray} \label{eq:f^1}
\epsilon_{ijk}\partial_i f^1_{jk} & = & -\epsilon_{ijk}\partial_i 
           (\theta_{mn}\partial_m A^0_j\partial_n A^0_k)
            -\epsilon_{ijk}\theta_{nm}\partial_n A^0_k\partial_m F^0_{ij}
                                      \nonumber\\
 & = & -\epsilon_{ijk}\theta_{mn}\Big(\partial_m \partial_i A^0_j
                  \partial_n A^0_k
            +\partial_mA^0_j\partial_n\partial_i A^0_k   
            -\partial_n A^0_k\partial_m (\partial_i A_j^0-
                   \partial_j A_i^0)\Big).
\end{eqnarray}}

\vskip -0.5cm
\noindent The f\mbox{}irst and the third terms cancel each other; 
while shuf\mbox{}f\mbox{}ling
the indices of the fourth term ($i\to k$, $j\to i$, $k\to j$ and 
$m\leftrightarrow n$) makes it cancel the second one. So the right hand side of
(\ref{eq:f^1}) vanishes, {\it i.e.}, $d f^1=0$. This is a source-free 
equation with solution $f^1=0$ (because of the boundary condition at
inf\mbox{}inity),
therefore we f\mbox{}ind that $A^1$ is a pure gauge. We proceed to the 
second order
{\setlength\arraycolsep{2pt}
\begin{eqnarray} \label{eq:f^2}
\epsilon_{ijk}\partial_i f^2_{jk} & = & -\epsilon_{ijk}\partial_i
           \Big(\theta_{mn}(\partial_m A^0_j\partial_n A^1_k-
                            \partial_m A^0_k\partial_n A^1_j)\Big) \nonumber\\
 & & -\epsilon_{ijk}\theta_{nm}\partial_n A^1_k\partial_m F^0_{ij}
     -\epsilon_{ijk}\theta_{mn}\partial_m A^0_k\partial_n F^1_{ij}.
\end{eqnarray}}

\vskip -0.5cm
\noindent It's easy to show that most of the terms in (\ref{eq:f^2}) cancel
out by either shuf\mbox{}f\mbox{}ling the indices or directly 
setting $A^1$ to be zero.
The only non-trivial term comes from the $A^0$ part in $F^1$. More precisely
we are left with
\begin{equation}
\epsilon_{ijk}\partial_i f^2_{jk}=-\epsilon_{ijk}\theta_{mn}\theta_{pq}
            \partial_m A^0_k\partial_n (\partial_p A^0_i\partial_q A^0_j).
\end{equation} 
The term on the right hand side is just $-2\epsilon_{ijk}(\theta_{mn}
                  \theta_{pq}\partial_m A^0_k\partial_q A^0_j
                  \partial_n \partial_p A^0_i)$ with the part in brackets
symmetric in $j$ and $k$, so it vanishes identically. Once again, we are
led to the usual Bianchi identity $d f^2=0$, which shows $A^2$ is trivial
up to a gauge transformation. 
One may argue that $f^{1, 2}$ are closed from the previous def\mbox{}inition.
But we note $f^{1, 2}=dA^{1, 2}$ are only def\mbox{}ined \emph{locally},
just like in the commutative case $F^0=dA^0$ is not valid \emph{globally}
on $S^2$. {\it A priori}, it is not obvious the source terms due to $A^0$ in
(\ref{eq:f^1}) and (\ref{eq:f^2}) should vanish.

It is tempting to generalize the above result to higher order: the gauge
potential $A$ does not receive any corrections at all. Since the mere 
existence of a non-trivial $A$ is due to the monopole at origin, while it's
hard to see why a physical quantity def\mbox{}ined at a \emph{single} point
should be modif\mbox{}ied because of the spatial non-commutativity (this
is also the reason why we do not put $g\simeq g^0+g^1+g^2$ in (\ref{eq:DF})).
In consequence, the monopole mass still diverges. Explicitly, $|A^0|\sim 1/r$,
$|F^0|\sim 1/r^2$, $|F^1|\sim 1/r^4$ and $|F^2|=0$, so
\begin{equation}
m=\int |F|^2 \sim \int_0^{\infty} r^2dr\left| \frac{1}{r^2}+\frac{1}{r^4}
     \right|^2=\infty.
\end{equation}
While this result is disappointing, there exists an alternative way to 
address the monopole problem because of the self-interaction of photons
in non-commutative space~\cite{mh}. As already remarked, the singularity 
at origin is put in by hand and results
in the inf\mbox{}inite mass. Instead of introducing the singularity beforehand,
one can try to seek f\mbox{}inite energy solution to the equation of motion.
In other words, we are not looking for a point monopole, but an extended
f\mbox{}ield conf\mbox{}iguration just like in commutative non-Abelian
gauge theory. Now one faces the question as whether a Higgs f\mbox{}ield
should be introduced. In the brane picture, a non-trivial $\phi$ describes
the shape of the brane. But it's not clear whether the topological
charge depends on $\phi$. A detailed analysis is beyond the scope of this
short paper and may appear in a subsequent report.

There is the important question about gauge invariance. Due to 
the $\star$-product structure, the $U(1)$ f\mbox{}ield strength is not 
gauge invariant any more
{\setlength\arraycolsep{2pt}
\begin{eqnarray}
\delta F & \simeq & [i\lambda, F]_{\star} \nonumber\\
 & \simeq & 0-\theta_{mn}\partial_m \lambda^0\partial_n F^0
            -(\theta_{mn}\partial_m \lambda^1\partial_n F^0
              +\theta_{mn}\partial_m \lambda^0\partial_n F^1),
\end{eqnarray}}

\vskip -0.5cm
\noindent where $\lambda=\lambda^0+\lambda^1+\cdots$ is the gauge parameter.
This makes it dif\mbox{}f\mbox{}icult to interpret what is the magnetic 
f\mbox{}ield generated by the monopole. For instance, consider an electron
moving in the monopole background. Even on the classical level, one has
to modify the Lorentz force formula $\vec f =q\vec v\times \vec B$ in order
to make sure $\delta \vec f =0$ under $\delta B=\ast \delta F$ (or 
equivalently it imposes constraints on the possible form of gauge 
transformation). This situation dif\mbox{}fers too much from the 
familiar physical
facts, thus raises a basic question one bears in mind: what is the reality
of spacetime non-commutativity?

To achieve non-commutativity, one has to give a non-zero expectation value
to the Neveu-Schwarz tensor f\mbox{}ield, hence the associated
f\mbox{}ield energy (we do not consider the so-called \emph{transverse
non-commutativity}~\cite{ew} here). This adds to the total vacuum energy 
density $\rho_{\rm V}$ and 
therefore is related to the long-standing cosmological constant 
problem (for a review, see~\cite{w}). The non-commutativity scale is
determined by the background f\mbox{}ield $\theta =1/B$. If we assume, 
for a time, the vacuum energy comes solely from $B$ f\mbox{}ield, we can
calculate $\theta$ by using the observed value of $\rho_{\rm V}$.
Assuming the energy density of $B$ takes the same form as that of
magnetic f\mbox{}ield\footnote{One needs to be a little cautious: although $B$
resembles a magnetic f\mbox{}ield, it is a two-form \emph{potential}, not a
\emph{f\mbox{}ield strength} in string theory.}, we have
{\setlength\arraycolsep{2pt}
\begin{eqnarray} 
\rho_{\rm V} & = & \frac{1}{8\pi}B^2=\frac{1}{8\pi \theta^2},\\
             &   & \nonumber
\end{eqnarray}}

\vskip -0.5cm
\noindent which gives
\begin{equation}
\sqrt{\theta}=\left(\frac{1}{8\pi \rho_{\rm V}}\right)^{1/4}\simeq 
          \left(\frac{1}{8\pi \times 10^{-47}\,\,\mathrm{GeV}^4}\right)^{1/4}
          \simeq 5.0\times 10^{-3}\,\,\mathrm{cm}. 
\end{equation}
This numerical value is hardly believable as otherwise it 
would be verif\mbox{}ied
easily by experiment (note a portion of $\rho_{\rm V}$ only makes the
result worse). Of course the above simple calculation does not exclude
the possibility that spacetime may be non-commutative on the fundamental
level, since the cosmological constant itself has a 120 orders of magnitude
contrast between theoretical estimation and experimental observation!
Nevertheless, it indicates the nature of non-commutative spacetime is far
from clear. A plausible test bed of non-commutativity is superconductivity.
Unlike the monopoles, which largely remain to be theorists' fond toys,
the magnetic vortices have been observed in type II superconductors and 
the associated Abelian Higgs model has been well studied (see, 
for instance,~\cite{jt}). Non-commutativity could show its signature via
some modif\mbox{}ications to the conventional f\mbox{}lux lattice.

In conclusion, we studied Dirac monopole in non-commutative space and
found no non-trivial corrections. A more interesting solution may appear
in an alternative formulation of the problem. In the process, we raised
more questions than we could answer. While spacetime non-commutativity 
seems too remote from the real world, we have no doubts about this 
increasing abstraction in mathematics and physics as remarked in Dirac's
original paper~\cite{pd}.    

\vskip 1cm
\noindent {\Large \bf Acknowledgements}\\

\noindent We would like to thank Miao Li for bringing to our attention the 
paper~\cite{dl}. This work was supported in part by NSF Grant PHY9511632 
and the Robert A.~Welch Foundation.

\end{document}